\documentclass{aa}
\usepackage{graphics}
\usepackage{psfig}
\usepackage{times} 

\begin{document}

   \thesaurus{11.06.1;	 12.03.3;   12.04.1}

\title{Observational tests of the $\delta_{\rm c}$-$M_{\rm vir}$
relation in hierarchical clustering models}

\author{Ling-Mei Cheng\inst{1,2} and Xiang-Ping Wu\inst{1,2}} 

   \institute{National Astronomical Observatories,
              Chinese Academy of Sciences, Beijing 100012, China
\and 
	Department of Astronomy, Beijing Normal University,
        Beijing 100875, China}

\offprints{L.-M. Cheng}
\mail{clm@class2.bao.ac.cn}

   \date{Received 16 March, 2001; accepted 5 April, 2001}

\titlerunning{Observational tests of the $\delta_{\rm c}$-$M_{\rm vir}$ 
              relation}
\authorrunning{Cheng \& Wu}

\maketitle

\begin{abstract}
Observational determinations of the correlation between the characteristic 
density $\delta_{\rm c}$ and the virial mass $M_{\rm vir}$ of dark halos 
constitute a critical test for models of hierarchical structure formation.
Using the dynamical properties of dark halos reconstructed from 
the galaxy distributions in massive systems
(groups/clusters) and the rotation curves in less massive systems
(dwarf, low surface brightness and spiral galaxies) drawn from the literature, 
we confirm the existence of the $\delta_{\rm c}$-$M_{\rm vir}$ relation
over a broad mass range from $10^{10}M_{\odot}$ to $10^{15}M_{\odot}$,
which is in gross consistency with the prediction of a flat cosmological
model with $\Omega_{\rm M}=0.3$ and $\Omega_{\Lambda}=0.7$.
It is pointed out that previous analyses based on
the measurements of X-ray emitting gas and the hydrostatic equilibrium 
hypothesis, which claimed a shallower scale-free spectrum of $n\approx-1$
for initial density fluctuations, 
may suffer from nongravitational heating influence especially in 
low-mass systems.
\keywords{cosmology: observations --- dark matter --- 
          galaxies: formation}
\end{abstract}

%
%  14.Sep.'90: Demo-Vs.
%________________________________________________________________

\section{Introduction}

In hierarchical clustering models, low-mass halos (e.g. galaxies)
are denser and have higher collapse redshift than massive halos
(e.g. clusters), while the latter form  by the gravitational 
aggregation of individual low-mass objects. 
For a halo density profile of the form 
$\rho_{\rm DM}(r)=\delta_{\rm c} \rho_{\rm c} \overline{\rho}(r/r_{\rm s})$
where $\rho_{\rm c}$ is the critical density of the universe, 
$\delta_{\rm c}$ and $r_{\rm s}$ are the characteristic density and
length, respectively, the above scenario implies that
the characteristic density $\delta_{\rm c}$ of virialized dark halos
should correlate with virial mass $M_{\rm vir}$.  Indeed, 
the existence of such a $\delta_{\rm c}$-$M_{\rm vir}$ relation has been 
confirmed by a number of high-resolution $N$-body simulations 
in various cosmological models 
(SCDM, LCDM and OCDM) with different power spectra (e.g. 
Navarro et al. 1997; Salvador-Sol\'e et al. 1998).

On the observational side, several attempts have been recently made to
test the $\delta_{\rm c}$-$M_{\rm vir}$ relation and hence, to
constrain the scale-free power spectrum ($P(k)\propto k^n$)
for primordial density fluctuations
on relevant scales. By fitting the X-ray surface 
brightness profiles of clusters predicted by the universal density
profile (Navarro et al. 1995; NFW) to the ones
for 63 rich clusters observed with ROSAT, Wu \& Xue (2000) found 
that $\delta_{\rm c}\propto M_{\rm vir}^{-1.2}$,
indicating $n\approx-0.7$. This result was subsequently confirmed by 
Sato et al. (2000)
based on an analysis of the mass profiles of 83 ASCA X-ray objects
over a wider mass range from $M_{\rm vir}=10^{12}M_{\odot}$ to
$10^{15}M_{\odot}$, which gives  $n=-1.2\pm0.3$ ($90\%$ confidence level). 
It appears that while the presence of good correlation between 
$\delta_{\rm c}$ and $M_{\rm vir}$ has been observationally established, the
resulting power spectra are somewhat shallower than the 
value ($n\approx-2)$ expected for typical CDM, scale-free cosmologies
on cluster scales. 
Possible reasons for this conflict have been outlined in Mahdavi (1999)
and Wu \& Xue (2000).  Basically, the above observational tests of 
the $\delta_{\rm c}$-$M_{\rm vir}$ relation based on X-ray observations of
the intragroup/intracluster
gas may suffer from the influences of preheating, cooling flows, 
non-thermal pressure and hydrostatic equilibrium hypothesis. 
Among these, preheating by
supernovae and AGNs in the early phase of structure formation may lead 
the intragroup/intracluster gas to extend out to larger radii
(e.g. Kaiser 1991; David et al. 1991; Ponman et al.  
1999; Bower et al. 2001), which can significantly affect our mass estimates 
from the X-ray observed surface brightness profiles especially
for galaxies and groups when incorporated with the hydrostatic equilibrium
hypothesis.

Unlike intracluster gas, galaxies in groups and clusters are essentially 
unaffected by the presence of preheating, cooling flows and non-thermal 
pressure and can, therefore, be regarded as better tracers of 
underlying gravitational potentials. In less massive systems such as 
the dark matter dominated dwarf galaxies or low surface brightness galaxies,
one is able to recover the dark matter profiles from the well-measured 
rotation curves (e.g. Carignan \& Beaulieu 1989). 
Dynamical properties of dark halos derived from the distributions of 
galaxies in groups/clusters and the rotation curves of dwarf, low
surface brightness and spiral galaxies, which spans several decades in mass,
have already been available in the literature. It thus becomes possible 
and also timely to have a close
examination of the $\delta_{\rm c}$-$M_{\rm vir}$ relation motivated by
hierarchical clustering models. We attempt to fulfill the task in
this paper, and compare our new determination with the results given
by X-ray observations and theoretical expectations.
Throughout this paper, we
assume $H_0=100$ km s$^{-1}$ Mpc$^{-1}$ and 
a flat cosmological model of $\Omega_{\rm M}=0.3$ and
$\Omega_{\Lambda}=0.7$. The shape of the $\delta_{\rm c}$-$M_{\rm vir}$ 
relation is roughly unaffected by this choice 
(Wu \& Xue 2000; Sato et al. 2000).

\section{Sample}

We model the virialized dark halos by the NFW universal density profile:
$\rho_{\rm DM}(r)=
\delta_{\rm c} \rho_{\rm c}/[(r/r_{\rm s})(1+r/r_{\rm s})^2]$,
where the critical density of the universe reads 
$\rho_{\rm c}=(3H_0^2/8\pi G)E^2(z)$ and 
$E^2(z)=(1+z)^2\{1+z\Omega_{\rm M}+[(1+z)^{-2}-1]\Omega_{\Lambda}\}$. 
The virial mass $M_{\rm vir}$ is defined by
$M_{\rm vir}=4\pi r_{\rm vir}^3 \Delta_{\rm c} \rho_{\rm c}/3$,
so that 
$\delta_{\rm c}=(\Delta_{\rm c}/3)\{c^3/[\ln(1+c)-c/(1+c)]\}$,
where $\Delta_{\rm c}$ is the overdensity of dark matter with respect to
the critical value $\rho_{\rm c}$ and can be approximated by 
$\Delta_{\rm c}=178\Omega_{\rm M}(z)^{0.45}$ (Eke et al. 1998), and 
$c=r_{\rm vir}/r_{\rm s}$ is called the concentration parameter.

Assuming that galaxies in groups and clusters trace mass, we are able
to fix the free parameter $r_{\rm s}$ by straightforwardly fitting 
the observed surface number density profiles of galaxies to 
the projection of the NFW profile. In order to obtain 
$\delta_{\rm c}$, we also need to have an independent estimate of 
the virial radius $r_{\rm vir}$ or the concentration parameter $c$.
An approximate approach to the problem is to employ virial theorem: 
$M_{\rm vir}=3\sigma_{\rm p}^2 r_{\rm vir}/G$, 
which yields $r_{\rm vir}=(6/\Delta_{\rm c})^{1/2}[\sigma_{\rm p}/H_0E(z)]$,
where  $\sigma_{\rm p}$ 
is the global line-of-sight velocity dispersion. 
However, this expression holds true only for clusters with 
an isotropic distribution of galaxies, and their 
velocity dispersion can be reliably measured out to virial radii.
Recall that the velocity dispersion of galaxies varies with radius 
if galaxies are assumed to trace mass (Cole \& Lacey 1996).
We have thus checked the
accuracy of the estimated virial radii using a sample of 158 nearby clusters
of Girardi et al. (1998), in which the virial radii have been given 
by dynamical analysis, and found that there is good agreement between
the two methods.  This technique has been applied to an ensemble of 
rich systems of galaxies by CNOC (Carlberg et al. 1997), and
poor systems of galaxies (PSG) by Mahdavi et al. (1999). 
Their averaged, best-fit parameters are summarized in Table 1. For PSG
we adopt the results of the sample D [PSG(D)] and eight groups with 
$\sigma_{\rm p}>350$ km s$^{-1}$ [PSG(H)] but exclude the result for the nine
groups with $\sigma_{\rm p}<350$ km s$^{-1}$ due to the failure of the NFW fit.
The third data set is taken from the ENACS cluster sample 
fitted by Adami et al. (1998). These authors used a similar technique but 
a pseudo NFW profile. We adopt their average value of $r_{\rm s}=0.26$ Mpc 
among the 41 clusters whose velocity dispersions are observationally 
determined. The recent work based on a sample of 77 composite ENACS clusters
gives the similar result,  $r_{\rm s}=0.318$ Mpc (Adami et al. 2001).

For less massive systems, the two free parameters in the NFW profile can be 
determined by utilizing the high-quality rotation curves. van den Bosch
\& Swaters (2001) have recently obtained the concentration parameters 
$c_{200}$ and the circular velocities $V_{200}$ at $r_{200}$ for a sample of 
20 dwarf galaxies by properly subtracting the contributions of
the thin gas disk and the thick stellar disk to the rotation curves.
We take the mean values of their best-fit $c_{200}$ 
and $V_{200}$ for the 15 dwarf galaxies (DWG) that satisfy $c_{200}>1$ 
and convert $c_{200}$ and $r_{200}$ into $c$ and $r_{\rm vir}$ in terms of 
the NFW profile and our definitions of $c$ and $r_{\rm vir}$. 
The major uncertainty in their fittings 
arises probably from the constant mass-to-light ratio $M/L$ assumed for
the stellar disk components. In order to demonstrate this uncertainty,
we adopt both the result for  $M/L=1(M/L)_{\odot}$ (DWG1)
and the one for $M/L=0$ (DWG0). 
Another sample we have used is the 9 low-luminosity disk galaxies (DIG) 
studied by Borriello \& Salucci (2001). These authors have worked out
the $c$ and $r_{\rm s}$ parameters  based on the fitting of the
rotation curves to the stellar disk plus dark halo models, in which
the mass-to-light ratio $M/L$ for the stellar disk is treated as 
a free parameter.  However, they found that good fits in DIG sample are 
obtained only for unreasonably large virial velocities and masses.
To overcome the inadequacy, they imposed an upper limit of 
$2\times10^{12}M_{\odot}$ on $M_{\rm vir}$ in their fitting. 
We use their average values of $c$ and $r_{\rm s}$ after 
the cosmological model correction is properly made.
We have also performed the fitting of 
the rotation curves predicted by the NFW halo plus thin stellar disk model to
a sample of 30 spiral galaxies in the Ursa Major (UMA) cluster analyzed
by Sanders \& Verheijen (1999). However, the models are not well constrained
for majority of galaxies due to the sparse data points of the rotation 
curve if the mass-to-light ratio of stellar components is 
treated as a free parameter and no upper limit is imposed on
the virial masses of dark halos.  We have only used 
the results of 13 galaxies that satisfy $c>1$, and their best-fit 
parameters are listed in Table 1.

As a comparison, we turn to the best-fit values of $r_{\rm s}$ and
$c$ from the X-ray measurements of 63 rich clusters observed with
ROSAT (Wu \& Xue 2000) and 83 objects observed with ASCA (Sato et al. 2000).
While these two groups both assumed hydrostatic equilibrium for 
X-ray emitting gas, they used very different methods in the 
determinations of $r_{\rm s}$ and $c$: Wu \& Xue (2000) directly 
fitted the theoretically predicted X-ray surface brightness profiles 
$S_{\rm X}(r)$ from the NFW profile via isothermality 
(Makino et al. 1998) to the ROSAT observed $S_{\rm X}(r)$;
Sato et al. (2000) first derived the mass distributions
of groups/clusters from the X-ray observed surface brightness profiles 
characterized by conventional $\beta$ model and the temperature 
profiles. They then calculated $\delta_{\rm c}$ and $r_{\rm s}$ by
fitting the NFW model to the mass profiles. For our analysis below,
we use 3 and 4 bins according to virial mass
for the ROSAT and ASCA samples, respectively (see Table 1).

\begin{table*}
 \vskip 0.2truein
 \scriptsize
 \begin{center}
 \caption{Sample}
 \begin{tabular}{lllllllllc}
 \hline
 & & & & & & & & & \\
sample &  No & \multicolumn{1}{c}{redshift} & $\sigma_{\rm p}$ (km s$^{-1}$) &
$kT$ (keV) & $r_{\rm s}$ (Mpc) & \multicolumn{1}{c}{$c$}  &
$M_{\rm vir}$ ($10^{14}M_{\odot}$) & $\delta_{\rm c}$ ($10^4$) & ref\\
 & & & & & & & & & \\
CNOC &  14  & $0.313\pm0.120$ & $972\pm48$ & \multicolumn{1}{c}{...} &  
	$0.48^{+0.32}_{-0.26}$ & $3.70^{+3.99}_{-1.38}$ &
        $11.76^{+1.82}_{-1.65}$  & $0.29^{+1.25}_{-0.18}$ & 1 \\
ENACS&  41  & $0.073\pm0.027$ & $617\pm38$ & \multicolumn{1}{c}{...} &
	$0.26\pm0.03$ & $5.42^{+1.19}_{-0.92}$ & 
        $3.70^{+0.73}_{-0.65}$ & $0.58^{+0.34}_{-0.20}$ & 2 \\
PSG(H)  &  8   & $0.025\pm0.000$ & $460\pm38$  & \multicolumn{1}{c}{...} & 
        $0.17\pm0.05$ & $6.00\pm1.70$  & 
        $1.60^{+0.43}_{-0.37}$ & $0.70^{+0.56}_{-0.37}$ & 3 \\
PSG(D)  &  5   & $0.024\pm0.000$ & $392\pm48$  & \multicolumn{1}{c}{...} & 
        $0.117\pm0.043$ &   $6.70\pm2.40$  & 
        $0.99^{+0.41}_{-0.32}$ & $0.91^{+0.98}_{-0.58}$ & 3 \\
UMA     & 13    & $0.004$ & \multicolumn{1}{c}{...} & \multicolumn{1}{c}{...} &
        $0.022\pm0.002$ &  $9.71\pm0.45$  & 
        $8.46^{+0.09}_{-0.09}\times10^{-3}$ & $1.97^{+0.48}_{-0.39}$ & 5 \\
DIG     & 9    & $\sim0$ & \multicolumn{1}{c}{...} & \multicolumn{1}{c}{...} &
        $0.023\pm0.002$ &  $7.37\pm0.62$  & 
        $5.61^{+4.16}_{-2.57}\times10^{-3}$ & $1.11^{+0.24}_{-0.21}$ & 4 \\
DWG1     & 15   & $\sim0$ & \multicolumn{1}{c}{...} & \multicolumn{1}{c}{...}& 
	$0.006\pm0.002$ & $15.0\pm2.3$ &
        $1.03^{+2.16}_{-0.78}\times10^{-3}$ & $6.33^{+2.81}_{-2.19}$ & 6 \\
DWG0     & 15   & $\sim0$ & \multicolumn{1}{c}{...} & \multicolumn{1}{c}{...}& 
	$0.004\pm0.001$ & $21.2\pm2.9$ &
        $0.74^{+1.32}_{-0.53}\times10^{-3}$ & $15.4^{+6.1}_{-4.9}$ & 6 \\
 & & & & & & & & & \\
\hline
 & & & & & & & & & \\
ROSAT&  20  & $0.139\pm0.042$ & \multicolumn{1}{c}{...}  & $9.23\pm0.57$ &
	$1.78\pm0.21$  & $3.44\pm1.81$ &
	$15.6^{+8.8}_{-5.6}$ & $0.17^{+0.37}_{-0.12}$ & 7 \\
     &	25  & $0.107\pm0.035$ & \multicolumn{1}{c}{...}  & $6.83\pm0.73$ &
	$0.66\pm0.02$  & $6.87\pm3.68$ &
	$5.83^{+2.25}_{-1.62}$ & $0.75^{+2.16}_{-0.56}$ & 7 \\
     &  18  & $0.069\pm0.040$ & \multicolumn{1}{c}{...}  & $3.98\pm0.23$ &
	$0.24\pm0.02$ & $12.2\pm7.3$ &
	$1.69^{+1.33}_{-0.75}$ & $2.92^{+7.29}_{-2.08}$ & 7 \\
ASCA &  30  & $0.178\pm0.037$ & \multicolumn{1}{c}{...}  & $7.51\pm0.72$ & 
	$0.281\pm0.042$  & $5.50\pm0.96$ &
	$5.48\pm0.62$ & $0.65^{+0.29}_{-0.23}$ & 8 \\
     &  30  & $0.120\pm0.050$ & \multicolumn{1}{c}{...}  & $5.25\pm0.76$ & 
	$0.155\pm0.037$  & $8.80\pm1.63$ &
	$2.36\pm0.31$ & $1.88^{+0.97}_{-0.73}$ & 8 \\
     &  10  & $0.024\pm0.013$ & \multicolumn{1}{c}{...}  & $2.72\pm1.37$ & 
	$0.043\pm0.009$  & $14.8\pm2.5$ &
	$0.50\pm0.13$ & $6.23^{+3.08}_{-2.35}$ & 8 \\
     & 	13  & $0.012\pm0.006$ & \multicolumn{1}{c}{...}  & $0.95\pm0.23$ & 
	$0.014\pm0.005$  & $35.8\pm10.2$ &
	$0.11\pm0.03$ & $60.9^{+57.9}_{-35.7}$ & 8 \\
 & & & & & & & & & \\
 \hline
 \end{tabular}
 \end{center}

\parbox{7.in}{References: 
(1)Carlberg et al. (1997);
(2)Adami et al. (1998); 
(3)Mahdavi et al. (1999);
(4)Borriello \& Salucci (2001);
(5)Sanders \& Verheijen (1998); 
(6)van den Bosch \& Swaters (2001);
(7)Wu \& Xue (2000); 
(8)Sato et al. (2000).
}
  \end{table*}

\section{Analysis and results}

Fig.1 shows the $\delta_{\rm c}$-$M_{\rm vir}$ relation
over a broad mass range from $10^{10}M_{\odot}$ to $10^{15}M_{\odot}$, 
in which the data points from X-ray emitting gas and 
those from distributions and rotation curves of galaxies 
are clearly marked. 
All the quoted  error bars are $95\%$ confidence limits.
Also plotted in Fig.1 is the corresponding $c$-$M_{\rm vir}$ relation
for the purpose of easier comparison with numerical results 
(e.g. Navarro et al. 1997).
It can be seen that the two data sets (galaxies and X-ray gas) 
are essentially consistent with 
each other on large mass scales  $M_{\rm vir}>10^{14}M_{\odot}$,
but differ significantly at low-mass end. The best-fit
relations based on the X-ray sample alone are
%1,2
\begin{eqnarray}
\delta_{\rm c}=10^{19.44\pm2.65}(M_{\rm vir}/M_{\odot})^{-1.06\pm0.17};\\
c=10^{7.34\pm1.10}(M_{\rm vir}/M_{\odot})^{-0.45\pm0.07}.
\end{eqnarray}

    \begin{figure} 
	\psfig{figure=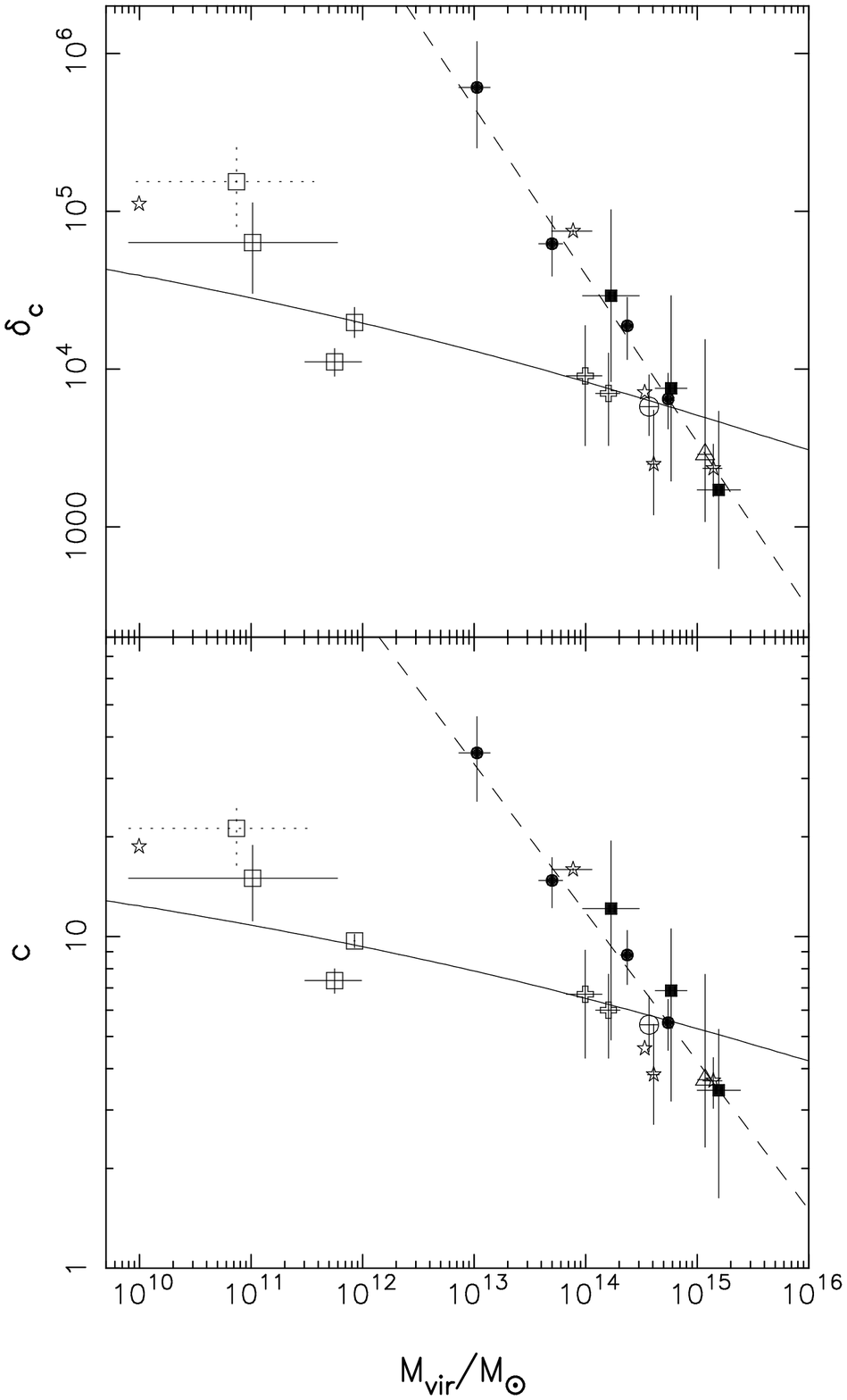,width=98mm,bbllx=80pt,bblly=70pt,bburx=540pt,bbury=750pt,clip=,angle=0}
	\caption{The observationally determined 
$\delta_{\rm c}$--$M_{\rm vir}$ (upper panel) and $c$-$M_{\rm vir}$
(lower panel) relations for two samples:
(1)galaxies -- CNOC (open triangle), ENACS (open circle), PSG (open cross), 
DWG0, DWG1, DIG and UMA (open squares from left to right); 
(2)X-ray gas -- ROSAT (fitted square) and ASCA (filled circle). 
The asterisks from left to right denote the results for five 
individual objects:  
DDO 154 from the measurement of HI rotation curve (Carignan \& Purton 1998;
Burkert \& Silk 1999), the Hydra A cluster observed 
recently with Chandra (David et al. 2000),
the poor cluster AWM 7 from the NFW fit to the radial profiles of
galaxies and their velocity dispersion (Koranyi \& Geller 2000), 
the nearest cluster Virgo from a combined analysis of optical
and X-ray observations (McLaughlin 1998), 
and the cluster Abell 576 from the measurement
of galaxy distribution out to the infall region (Rines et al. 2000).
Dashed line is the best-fit relation to the X-ray data. 
Solid line is the theoretical expectation for
a flat cosmological model of $\Omega_M=0.3$ and $\Omega_{\Lambda}=0.7$
with $\Gamma=0.25$, $\sigma_8=0.85$ and $n=1$.}
\label{fig1}
   \end{figure}

In order to compare with theoretical predictions in hierarchical 
clustering models, 
we compute the collapse redshift $z_{\rm coll}$ for halo of
mass $M_{\rm vir}$ identified at redshift $z=0$
in terms of the extended Press-Schechter formalism (Lacey \& Cole 1993;
Navarro et al. 1997):
%3
\begin{equation}
\delta_{\rm crit}(z_{\rm coll})=\delta_{\rm crit}(0)
      +0.477 \sqrt{2[\sigma^2(fM_{\rm vir})-\sigma^2(M_{\rm vir})]}, 
\end{equation}  
where $f\approx0.01$, $\delta_{\rm crit}(z)=
\delta_{\rm crit}(0)/D(z,\Omega_{\rm M},\Omega_{\Lambda})$ 
is the density threshold for spherical collapse at $z$ with
$D$ being the linear grow factor and $\delta_{\rm crit}(0)\approx1.69$,
and $\sigma(M_{\rm vir})$ is the rms linear density fluctuation
at mass scale $M_{\rm vir}=(4\pi/3)\rho_{\rm c}\Omega_{\rm M}R^3$, 
which is related to the fluctuation power
spectrum by 
%4
\begin{equation}
\sigma^2(M_{\rm vir})=
\frac{1}{2\pi^2}\int_{0}^{\infty} dk k^2P(k) W^2(kR). 
\end{equation}  
We use the standard top hat window function and parameterize the fluctuation 
power spectrum  by $P(k)=Ak^nT^2(k)$, in which the transfer function
$T(k)$ is taken from an adiabatic CDM model given by Bardeen et al. (1986)
for the shape parameter $\Gamma=0.25$ and the Harrison-Zeldovich spectrum 
$n=1$.  Furthermore, we 
assume that the rms fluctuation amplitude within a sphere of $R=8$ Mpc 
is $\sigma_8=0.85$.

In terms of the prescription of Navarro et al. (1997), the characteristic 
density $\delta_{\rm c}$ of a virialized dark matter halo 
$M_{\rm vir}$ is proportional to the mean density of the 
universe at the corresponding collapse redshift $z_{\rm coll}$
%5
\begin{equation}
\delta_{\rm c}=C\Omega_{\rm M}[1+z_{\rm coll}(M_{\rm vir})]^3,
\end{equation}
where $C$ is  the proportionality constant and can be approximately taken 
to be $C=3000$ for our choice of cosmological parameters.
In particular, for a power-law spectrum of primordial
density fluctuations, the above expression reduces to   
$\delta_{\rm c}\propto M_{\rm vir}^{-(n+3)/2}$ when combined with
equation (3).  Therefore, the best-fit $\delta_{\rm c}$-$M_{\rm vir}$ 
relation equation (1) from the X-ray data indicates $n=-0.88\pm0.24$, which
is simply the combined result of Wu \& Xue (2000) and Sato et al. (2000). 
The theoretically predicted 
$\delta_{\rm c}$-$M_{\rm vir}$ and $c$-$M_{\rm vir}$ relations
from equation (5) are illustrated in Fig.1.
It is immediate that the data points of the galaxy sample 
($\bar{z}<0.3$) are roughly consistent with the theoretical prediction.

\section{Discussion and conclusions}

Using the dynamical properties of dark halos reconstructed from 
galaxy distributions in massive systems
(groups/clusters) and rotation curves in less massive systems
(dwarf, low surface brightness and spiral galaxies) 
drawn from the literature, 
we have examined the $\delta_{\rm c}$-$M_{\rm vir}$ relation predicted by 
hierarchical clustering models. It turns out that
the observational data are in gross consistency with the theoretical 
prediction on a broad mass range from $10^{10}M_{\odot}$ to $10^{15}M_{\odot}$,
although the present analysis still suffers from the sparse data points 
especially on the low-mass scale of $M<10^{13}M_{\odot}$. 
Indeed, this last point reflects the difficulty in the current determinations 
of the dark matter profiles of galaxies from their rotation curves.
In fact, even if the high-quality rotation curves of HI disks can be traced 
out to large radii, the unknown mass-to-light ratio of the stellar components
can result in ambiguity regarding the determinations of $\delta_{\rm c}$
and $r_{\rm s}$ in the NFW dark matter profile 
(e.g. van den Bosch et al. 2000).
We have illustrated in Fig.1 the influence of $M/L$ on  
the $\delta_{\rm c}$-$M_{\rm vir}$ and $c$-$M_{\rm vir}$ relations 
for DWG sample. We cannot exclude the possibility that
the best-fit values of $\delta_{\rm c}$, $r_{\rm s}$ and $c$ in
other two samples (UMA and DIG) may 
still contain rather large uncertainties due to the unknown $M/L$
although $M/L$ is allowed to vary in the fitting processes. 
Recall that  good fits in DIG sample are 
obtained only for unreasonably large virial velocities and masses,
while majority of the galaxies in UMA sample simply failed in the fitting of
NFW profile. So, more accurate and reliable data of galactic systems will be 
needed to tighten the constraints on 
the $\delta_{\rm c}$-$M_{\rm vir}$ and $c$-$M_{\rm vir}$ relations
on low-mass scales. Alternatively, the assumption that 
galaxies trace mass has been often used in groups and clusters 
in order to extract the dynamical properties of dark halos from the
optically observed radial profiles of galaxies (e.g. CNOC,
ENACS and PSG). A quantitative estimate of the uncertainties 
in the determinations of the NFW profile from this oversimplification
of galaxy distributions should be made in the future.

The newly established 
$\delta_{\rm c}$-$M_{\rm vir}$ relation from `galaxies' differs
remarkably from previous findings based on X-ray
measurements of the hot diffuse gas contained in galaxies, groups and
clusters (Wu \& Xue 2000; Sato et al. 2000).
The latter results in a shallower power-law spectrum 
($n\approx-1$) than the expectations of typical CDM models on cluster scales.
It is possible that the conflict is associated with 
nongravitational heating of the X-ray gas in the early phase of structure 
formation. Such a speculation is supported by the good consistency of the
$\delta_{\rm c}$-$M_{\rm vir}$ (or $c$-$M_{\rm vir}$) 
relations between the two samples
(galaxies and gas) in massive systems ($M>10^{14}M_{\odot}$)
and the remarkable difference on low-mass scales.
Note that galaxies are less affected by the possible existence of preheating. 
This situation is similar to the recent
discovery of an entropy floor in the centers of groups and clusters 
(Ponman et al. 1999) due to the additional heating of the gas from
supernovae and/or AGNs. Smaller systems (groups and galaxies) have
lower entropies and hence are more affected than clusters. 
As a result, the present-day saturated configuration of hot/warm gas 
in less massive 
systems is shallower than the distributions of dark matter and galaxies.
Reconstruction of the dark matter profiles under the assumption of
hydrostatic equilibrium without the detailed information about the gas 
distribution and temperature at large radii may contain large uncertainties.  
A quantitative analysis will be needed to address 
whether the departure of the $\delta_{\rm c}$-$M_{\rm vir}$ 
relation detected in X-ray observations from those found by 
rotation curves and distributions of galaxies and 
predicted by typical CDM models arises from this speculation.

Essentially, it is hoped that 
the $\delta_{\rm c}$-$M_{\rm vir}$ and $c$-$M_{\rm vir}$
relations can be extended to
less massive systems below $10^{10}M_{\odot}$. This last point is potentially 
important for distinction between CDM and warm dark matter models
(Eke et al. 2000). Alternatively, the employment of the 
$\delta_{\rm c}$-$M_{\rm vir}$ relation established in the present work
can allow one to test the different prescriptions of 
halo formation time (e.g. Navarro et al. 1997; Salvador-Sol\'e et al. 1998; 
Eke et al. 2000; Bullock et al. 2001), and we 
will report the result elsewhere.

\begin{acknowledgements}
We thank Yan-Jie Xue for helpful discussion, Shinji Sato for 
sending us his X-ray catalog, and the referee, Cedric Lacey
for constructive suggestions that improved the presentation of
this work. This work was supported by 
the National Science Foundation of China, under Grant 19725311
and the Ministry of Science and Technology of China, under Grant 
NKBRSF G19990754.
\end{acknowledgements}

\end{document}